\documentclass[11pt]{article}
\usepackage{moriond,epsfig}

\usepackage{psfrag}

\def\mco{\multicolumn}

\def\be{\begin{equation}}
\def\ee{\end{equation}}
\def\bea{\begin{eqnarray}}
\def\eea{\end{eqnarray}}

\newcommand{\pbarnt}{\,\mbox{{\rm pb$^{-1}$}}}
\newcommand{\gev}{\,\mbox{GeV}}

\newcommand{\GeV}{\rm GeV}

\newcommand{\pb}{\rm pb}

\newcommand{\rpv}{\mbox{$R_p \!\!\!\!\!\! / \;\; $}}

\newcommand{\sell}{\tilde{e}_L}
\newcommand{\neu}{\tilde{\chi}_1^0}
\newcommand{\ev}{\,\mbox{eV}}

\begin{document}
\vspace*{4cm}
\title{SEARCHES FOR NEW PHYSICS IN $ep$ COLLISIONS AT HERA}

\author{ C. Schwanenberger}

\address{
(on behalf of the H1 and ZEUS collaborations)\\
Universit{\"a}t Bonn,\\
Nu{\ss}allee 12, 53115 Bonn, Germany
}

\maketitle\abstracts{
Recent results from searches for physics beyond the
Standard Model (SM)
in $e^\pm$-proton collisions at HERA at center-of-mass energies of 300
and 320 
GeV are presented. They were performed on a data sample collected in
the period $1994-2004$ by the H1 and ZEUS collaborations. The data have been
analysed searching for 
leptoquarks, light gravitinos in $R$-parity violating supersymmetric models and
magnetic monopoles. Results of a general search for new phenomena at high
transverse momentum and of a dedicated search for events with isolated leptons
and missing transverse momentum are also reported.
}

\section{Introduction}

At the HERA collider, electrons (positrons) and protons collide at a
center-of-mass energy of about $\sqrt{s}= 320$ GeV (300 GeV before
1998). From 1994-2000 (HERA I), integrated luminosities of about ${\cal L} =
113\ {\rm 
pb}^{-1}$ in $e^+ p$ and ${\cal L} = 17\ {\rm pb}^{-1}$ in $e^- p$
scattering were collected by each of the two experiments H1 and
ZEUS. These
data are used to search for new physics beyond the SM. Recent results on
searches for leptoquarks~\cite{leptoquarks} (section~\ref{sec:lq}), for
$R$-parity-violating (\rpv)
supersymmetry (SUSY)~\cite{gravitinos} (section~\ref{sec:grav}) and
magnetic monopoles~\cite{magmonopoles} (section~\ref{sec:mm}) are presented in
this 
contribution. 

A general search for new phenomena at high transverse momentum was recently
performed in a model independent framework by the H1
collaboration using the 1994-2000 $e^\pm p$ data sample~\cite{general} and was
repeated with the newest data taken after the HERA luminosity upgrade in
2003/2004 (HERA II) 
corresponding to ${\cal L} = 45 \pbarnt$ ($\sqrt{s}=320
\gev$)~\cite{highpt}. Results 
are presented in section~\ref{sec:general}.

The excess of events with isolated electrons or muons and missing transverse
momentum ($p_T$) observed in the period 1994-2000 by the H1
collaboration~\cite{isoleph1,isoleph1tau} has motivated a repeat of the same analysis on
the recent data taken in 2003/2004
corresponding to ${\cal L} = 53 \pbarnt$ ($\sqrt{s}=320 \gev$). In
section~\ref{sec:isolep} the event yields are compared with results obtained
by the ZEUS collaboration~\cite{isolepzeus1,isolepzeus2} performed on data
taken from 1994-2000 (${\cal L} = 130 \pbarnt$).
%where a slight excess of tau
%events compared to the SM expectation was seen. 

\section{Leptoquarks}\label{sec:lq}

The recent observations of neutrino oscillations have shown that lepton-flavor
violation (LFV) does occur in the neutrino sector. The LFV induced in the
charged-lepton sector due to neutrino oscillations cannot be measured at
existing colliders due to the low expected rate. However, there are many
extensions of the SM such as grand unified theories, SUSY, compositeness and
technicolor that predict possible $e \rightarrow \mu$ or $e \rightarrow \tau$
transitions at detectable rates.

%%%%%%%%%%%%%%%%%%begin%%figure%%%%%%%%%%%%%%%%%%%%%%%%%%%%%%%%%%%%%%%
\begin{figure}[hhh]
%\vspace{-1cm}
  \begin{center}
%   \psfrag{pi}[][][1.3][0]{ }
   {\epsfig{file=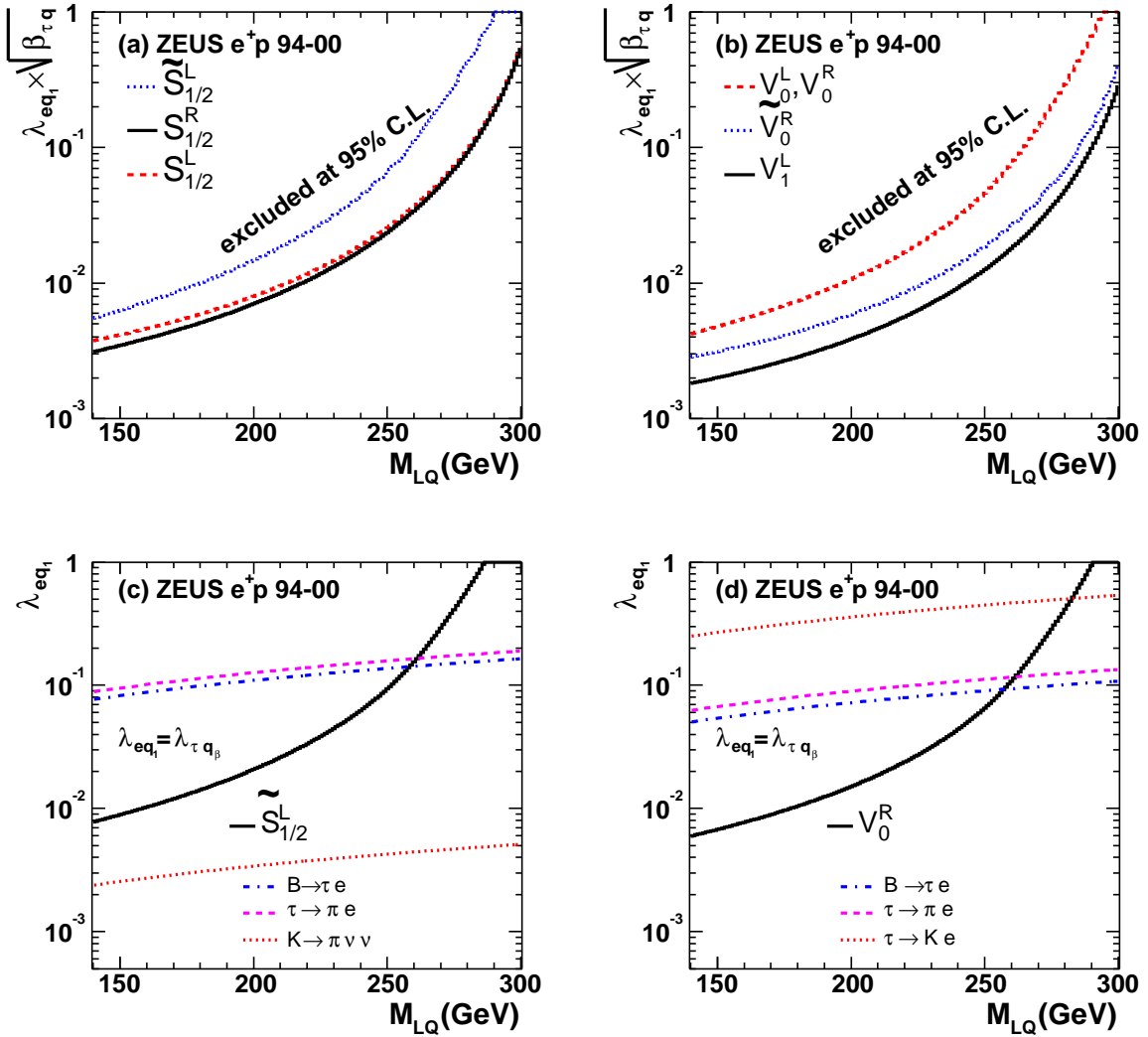,width=16cm}}
  \end{center}
%  \vspace{7.3cm}
%%
  \flushleft
  \caption{
Limits for $F=0$ LQs in the $\tau$ channel obtained from
  $e^+p$ collisions. The upper plots show $95 \%$ CL limits on
  $\lambda_{eq_1} \times \sqrt{\beta_{\tau q}}$ for (a) scalar and (b) vector
  LQs. In the lower plots, ZEUS limits on $\lambda_{eq_1}$ for a
  representative (c) scalar and (d) vector LQ are compared to the indirect
  constraints from low-energy experiments,
  assuming
  $\lambda_{eq_1} = \lambda_{\tau q}$.
}
  \label{fig:lq}
\end{figure}
%%%%%%%%%%%%%%%%%%%%%%%%%%%%%%end%%figure%%%%%%%%%%%%%%%%%%%%%%%%%%%%%

A search for LFV interactions $ep \rightarrow \mu X$
and $ep \rightarrow \tau X$ has been performed with the ZEUS detector using the
entire HERA I data sample~\cite{leptoquarks}. The presence of such processes,
which can be 
detected almost without background, would clearly be a signal of physics
beyond the SM. 
%This search is sensitive to all quark generations for LFV
%occurring between $e$ and $\mu$ or $\tau$.
No evidence for LFV was found and constraints were derived on LQs
that could mediate such 
interactions. The Buchm{\"u}ller-R{\"u}ckl-Wyler (BRW) LQ
model~\cite{brw} is used to set limits from the search. LQs are bosons that
carry both leptonic ($L$) and baryonic ($B$) numbers and have lepton-quark
Yukawa couplings. Their fermionic number ($F=3B+L$) can be $F=0$ or
$|F|=2$. Such bosons arise naturally in unified theories that arrange quarks
and leptons in common multiplets. A LQ that couples both to electrons and to
higher-generation leptons would induce LFV in $ep$ collisions through $s$- and
$u$-channel processes.

For LQ masses below $\sqrt{s}$, limits at $95\%$ confidence
level (CL) as a function of the mass were set on
$\lambda_{eq_1}\sqrt{\beta_{\ell q}}$, where $\lambda_{eq_1}$ is the coupling
of the LQ to an electron and a first-generation quark $q_1$, and $\beta_{\ell
  q}$ is the branching ratio of the LQ to the final-state lepton $\ell$ ($\mu$ 
or $\tau$) and a quark $q$. For a coupling constant of electromagnetic
strength ($\lambda_{eq_1}=\lambda_{\ell q}=0.3$), mass limits between 257 and
$299 \gev$ were set, depending on the LQ type. For $M_{LQ}=250 \gev$, upper
limits on $\lambda_{eq_1}\sqrt{\beta_{\mu q}}$
($\lambda_{eq_1}\sqrt{\beta_{\tau q}}$) in the range $0.010 - 0.12 (0.013 -
0.15)$ were set. 
As an example, in the $\tau$ channel upper limits
on $\lambda_{eq_1}\sqrt{\beta_{\tau q}}$ are shown in Fig.~\ref{fig:lq} for
$F=0$ scalar and vector LQs as a function of the LQ mass, assuming resonantly
produced LQs as described by the BRW model. In Figs.~\ref{fig:lq} (c) and (d),
the results are compared to constraints from rare
$\tau$, $B$ or $K$ decays~\cite{lowenergy}. ZEUS limits improve on low-energy
results in most 
cases.

For LQ masses much larger than $\sqrt{s}$, limits
were set on the four-fermion interaction term $\lambda_{eq_\alpha}
\lambda_{\ell q_\beta} / M^2_{LQ}$ for LQs that couple to an electron and a quark
$q_\alpha$ and to a lepton $\ell$ and a quark $q_\beta$, where $\alpha$ and
$\beta$ are quark generation indices. Some of the limits are also applicable
to LFV processes mediated by squarks in
\rpv\ SUSY models. In some cases, especially when a
higher-generation quark is involved and for the process $ep \rightarrow \tau
X$, the ZEUS limits are the most stringent to date.

\section{Light gravitinos in $R$-parity-violating SUSY}\label{sec:grav}

In Gauge Mediated Supersymmetry Breaking (GMSB) models, new
``messenger'' fields are introduced which couple to the source of
supersymmetry breaking. The breaking is then transmitted to the SM fields
and their superpartners by gauge interactions.
The gravitino, $\tilde{G}$, is the lightest
supersymmetric particle
(LSP) and can be as light as $10^{-3}\,\ev$.

An investigation of \rpv\ SUSY in
a GMSB scenario was performed by the H1 collaboration for the first time at
HERA~\cite{gravitinos}.
$R$-parity is a discrete multiplicative symmetry which can be written as
$R_p=(-1)^{3B+L+2S}$, where $B$ denotes the baryon number, $L$ the lepton
number and $S$ the spin of a particle.
The most general supersymmetric theory that is renormalisable and gauge
invariant with respect to the SM gauge group
contains \rpv\ Yukawa
couplings between the supersymmetric partner of the left-handed electron
$\tilde{e}_L$, a
left-handed
up-type quark $u_L^j$ and a right-handed down-type anti-quark
$\bar{d}_R^k$, where $j$ and $k$ denote generation indices. The
corresponding part of the Lagrangian reads
\begin{equation}
  {\cal L}_{\rpv} = - \lambda'_{1jk} \tilde{e}_L u_L^j \bar{d}_R^k + {\rm h.c.}
  \label{eq:rpv}
\end{equation}
At HERA, the presence of couplings $\lambda'_{1j1}$ and $\lambda'_{11k}$
could lead to resonant single neutralino production
in $e^+p$ and $e^-p$ collisions, respectively, via $t$-channel selectron
exchange, which is investigated here \footnote{
Resonant squark production in \rpv\
SUSY
has been investigated previously at HERA
in models in which the LSP is either a gaugino \cite{rpvhera} or a light
squark \cite{Aktas:2004tm}. Squark mass dependent limits
on various \rpv\ Yukawa couplings have been derived. In contrast, the
process considered in this analysis is completely independent of the squark
sector.}.
%A search for \rpv\ resonant
%single neutralino production $\neu$ via $t$-channel selectron exchange, $e^\pm q
%\rightarrow \neu q'$, is performed in $e^+p$ and $e^-p$ collisions.
It is assumed that the $\neu$ is
the next-to-lightest supersymmetric particle (NLSP) and that the decay $\neu
\rightarrow \gamma \tilde{G}$ occurs with an unobservably
small lifetime and dominates over \rpv\
neutralino decays. The resulting
experimental signature which was investigated in this analysis for the first
time at HERA is a photon, a jet originating from the
scattered quark and
missing transverse momentum due to the escaping gravitino. The main SM
background arises from radiative charged current (CC) deep inelastic
scattering (DIS) with a jet, a photon and a neutrino in the final state.

%
%%%%%%%%%%%%%%%%%%begin%%figure%%%%%%%%%%%%%%%%%%%%%%%%%%%%%%%%%%%%%%%
\begin{figure}[hhh]
\vspace{-3.5cm}
  \begin{center}
\begin{picture}(0,100)
         \put(-235,-15){\epsfig{file=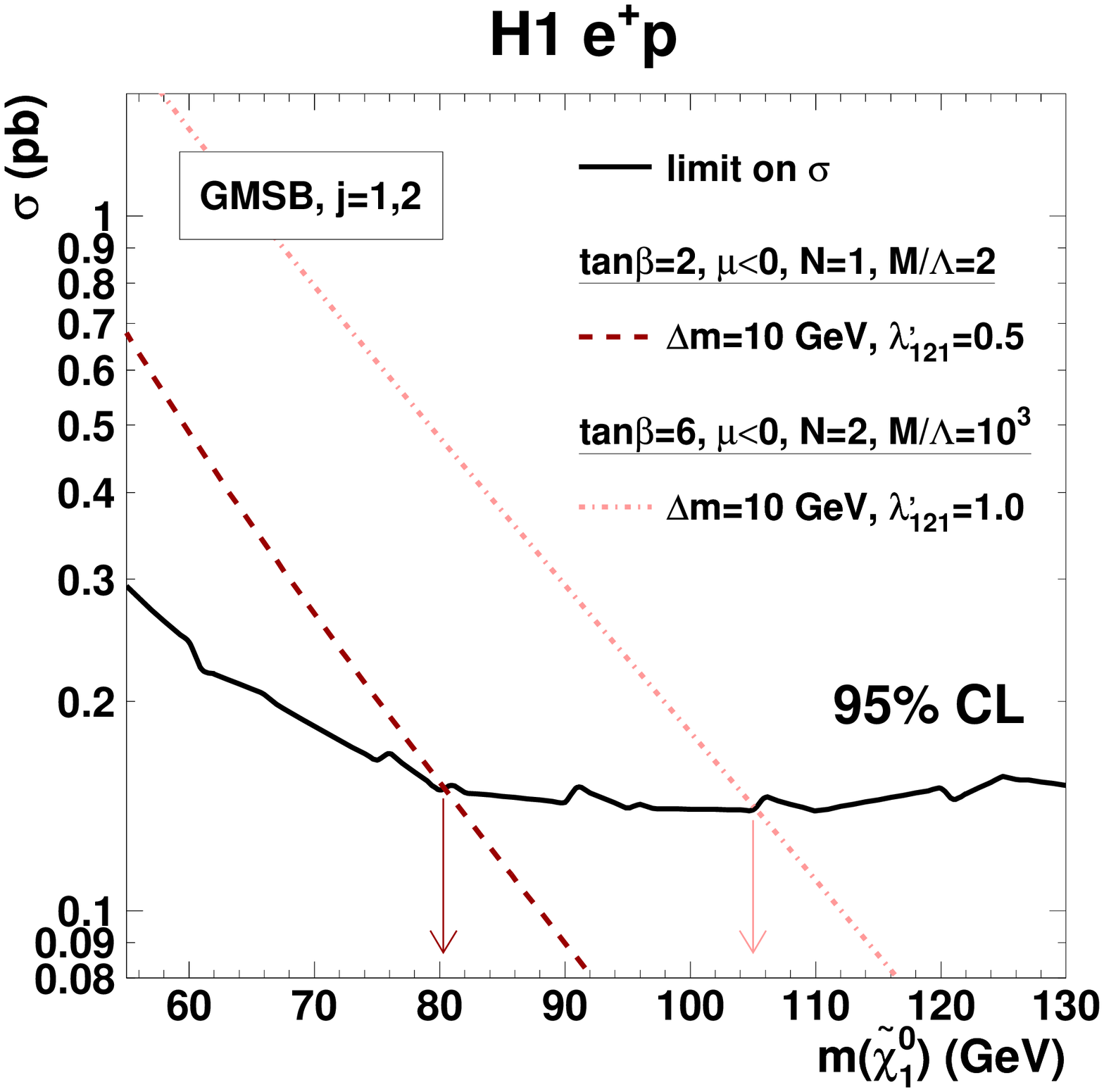,width=7.5cm,bbllx=25,bblly=660,bburx=575,bbury=130}}
         \put(0,-15){\epsfig{file=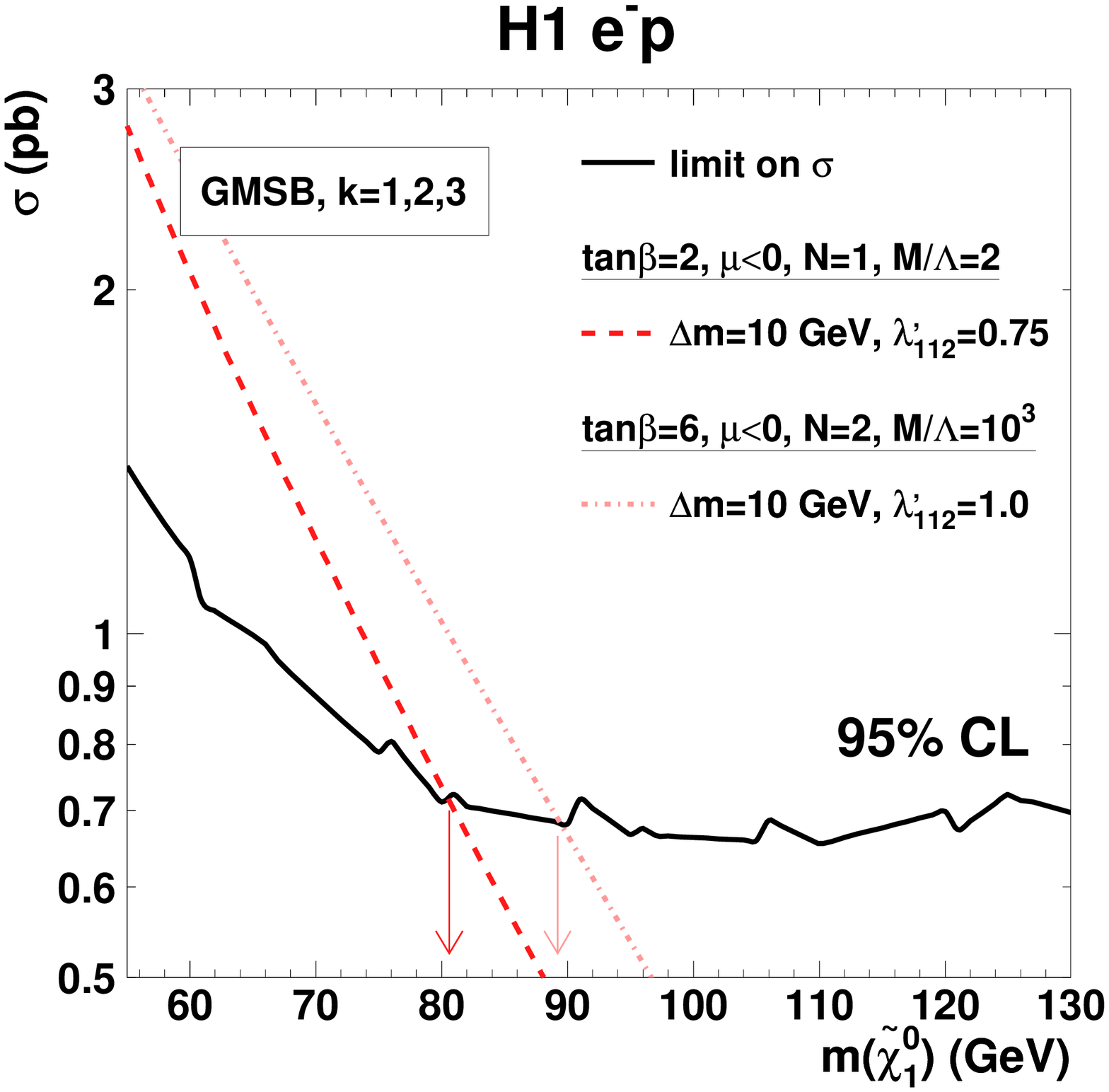,width=7.5cm,bbllx=25,bblly=660,bburx=575,bbury=130}}
\end{picture}
  \end{center}
  \vspace{7.5cm}
  \caption{Upper limit at the $95\,\%$ CL on the cross section as a function
    of the neutralino mass for example GMSB scenarios (solid lines). For
  comparison, the GMSB
    cross sections for different
    \rpv\ couplings $\lambda'_{121}$ and $\lambda'_{112}$
    are superimposed for a mass difference of $\Delta m =
  m({\tilde{e}_L})-m({\neu}) = 10 \,\GeV$ (dashed and dashed-dotted lines).}
  \label{fig:grav_xsec}
\end{figure}
%%%%%%%%%%%%%%%%%%%%%%%%%%%%%%end%%figure%%%%%%%%%%%%%%%%%%%%%%%%%%%%%
%
%%%%%%%%%%%%%%%%%%begin%%figure%%%%%%%%%%%%%%%%%%%%%%%%%%%%%%%%%%%%%%%
\begin{figure}[hhh]
\vspace{-4cm}
  \begin{center}
\begin{picture}(0,100)
         \put(-235,-15){\epsfig{file=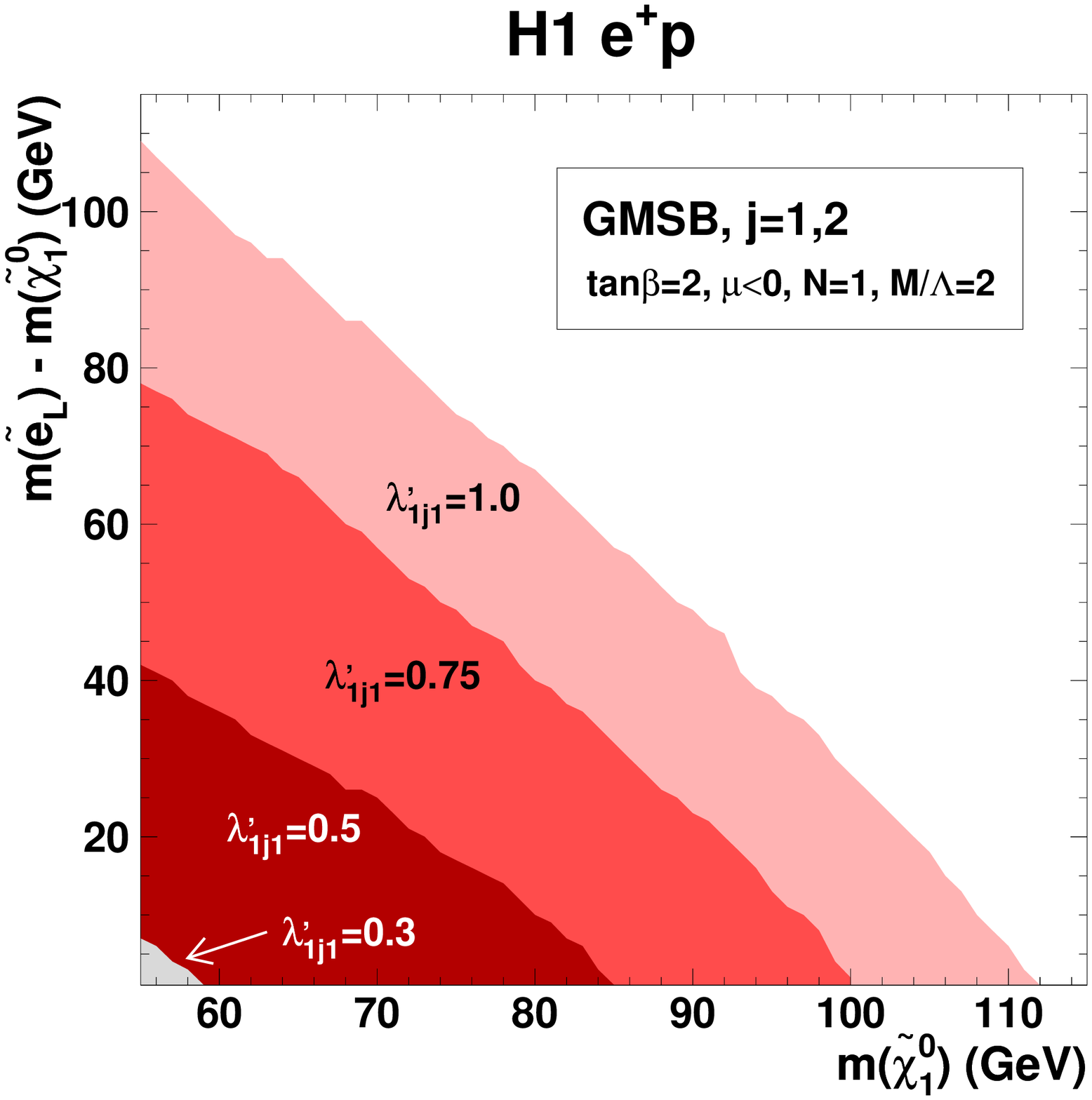,width=7.5cm,bbllx=25,bblly=660,bburx=575,bbury=130}}
         \put(0,-15){\epsfig{file=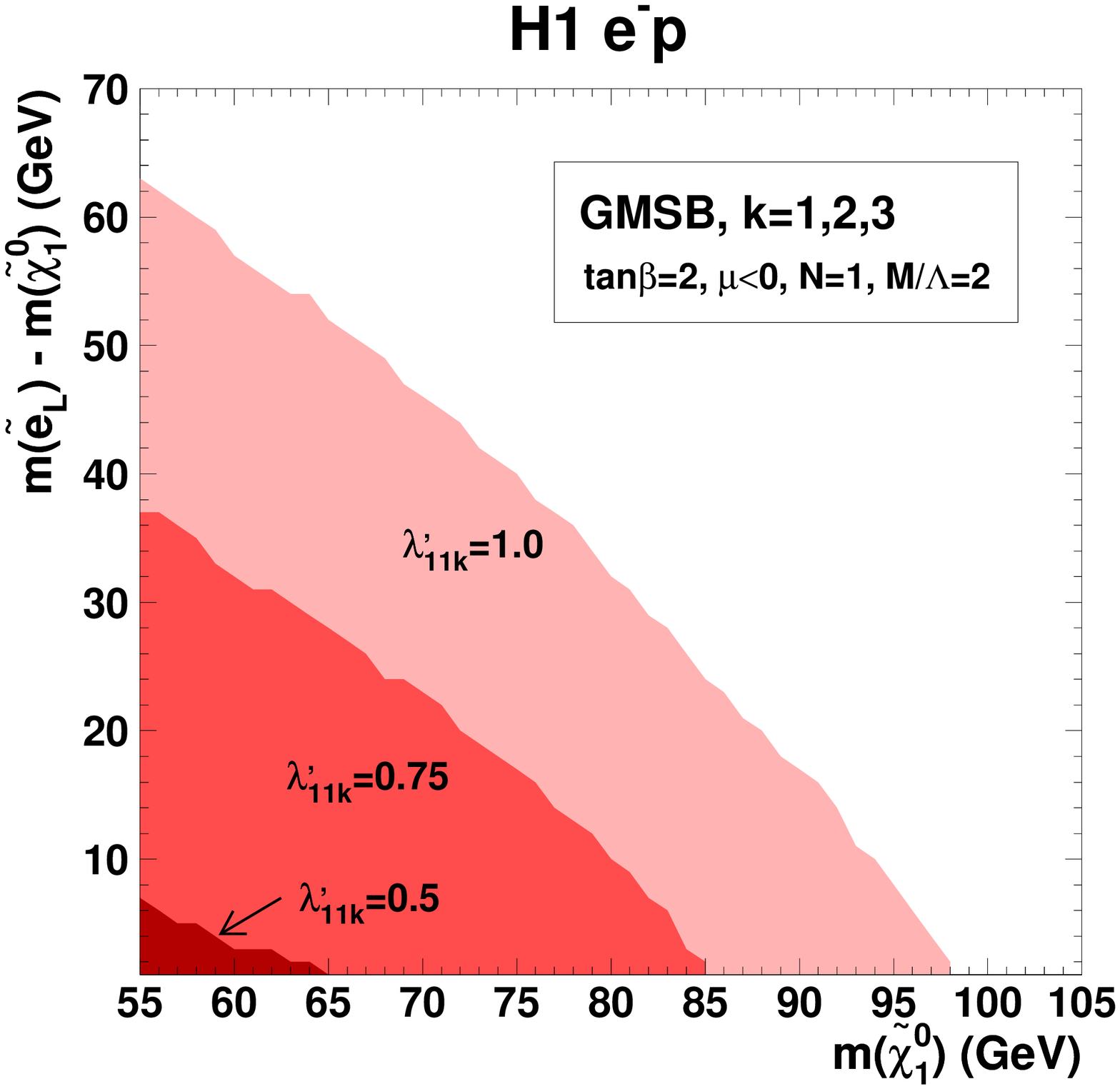,width=7.5cm,bbllx=25,bblly=660,bburx=575,bbury=130}}
\end{picture}
  \end{center}
  \vspace{7.5cm}
  \caption{Excluded regions at the $95\,\%$ CL in the
    $\Delta m = m({\tilde{e}_L})-m({\neu})$ and $m({\neu})$ plane
for various values of $\lambda'_{1j1}$ ($j=1,2$) and $\lambda'_{11k}$
($k=1,2,3$).
}
  \label{fig:grav_mass}
\end{figure}
%%%%%%%%%%%%%%%%%%%%%%%%%%%%%%end%%figure%%%%%%%%%%%%%%%%%%%%%%%%%%%%%

The data correspond to an integrated luminosity of $64.3\,\pb^{-1}$ for
$e^+p$ collisions recorded in 1999 and 2000 and $13.5\,\pb^{-1}$ for
$e^-p$ collisions recorded in 1998 and 1999.
The data analysis reveals no deviation from the SM.

Constraints on GMSB models are derived for different values
of the \rpv\ coupling.
In Fig.~\ref{fig:grav_xsec}, upper limits on
the cross sections are shown as a function of $m(\neu)$ for $e^+p$ (left) and
$e^-p$ (right)
collisions.
The limits become less stringent at low
neutralino masses due to the reduced signal detection efficiencies.
Typical GMSB cross sections
for different values of the couplings $\lambda'_{121}$ and $\lambda'_{112}$
are also shown for a mass difference $\Delta m =
m({\tilde{e}_L})-m({\neu}) = 10\,\GeV$.

In Fig.~\ref{fig:grav_mass}, excluded regions are
presented in the plane spanned by
$\Delta m $ and $m({\neu})$
using data from  $e^+p$ (left)
and $e^-p$ collisions (right) for various values of the respective \rpv\
coupling. 
For $\lambda'_{1j1} = 1.0$, the $e^+p$ results
exclude neutralino masses up to
$112\,\GeV$ for small $\Delta m$. For
large $\Delta m$ and small neutralino masses, selectron masses up to
$164\,\GeV$ 
are excluded. 
For masses $m(\neu)$ and $m(\sell)$ close
to $55\,\GeV$, $\lambda'_{1j1}$ Yukawa couplings of electromagnetic
strength are excluded.
In $e^-p$ collisions, for $\lambda'_{11k} = 1.0$, neutralino
masses up to 
$98\,\GeV$ for small $\Delta m$ and selectron masses up to $118\,\GeV$ for
large $\Delta m$ are ruled out.
These are the first constraints from HERA on SUSY
models which are independent of the squark sector.

\section{Magnetic Monopoles}\label{sec:mm}

One of the outstanding issues in modern physics is the question of the
existence of magnetic monopoles. Dirac showed that their existence leads
naturally to an explanation of electric charge
quantisation~\cite{dirac}. 
%Magnetic monopoles are also predicted from field
%theories which unify the fundamental forces.
The quantisation of the angular momentum of a system of an electron with
electric charge $e$ and a monopole with magnetic charge $g$ leads to Dirac's
charge quantisation condition $eg=n\hbar c/2$, where $\hbar$ is Planck's
constant divided by $2\pi$, $c$ is the speed of light and $n$ is an
integer. Within this approach, taking $n=1$ sets the theoretical minimum
magnetic charge which can be possessed by a particle (known as the Dirac
magnetic charge, $g_D$). However, if the elementary electric charge is
considered to be held by the down quark then the minimum value of this
fundamental magnetic charge will be three times larger. The value of the
fundamental magnetic charge could be even higher since the application 
of the Dirac argument to a particle
possessing both electric and magnetic charge restricts the values of $n$ to be
even. 

A direct search for magnetic monopoles produced in $e^+p$ collisions at HERA
has been made for the first time~\cite{magmonopoles}.
The beam pipe surrounding the interaction region in 1995-1997 was
investigated to look for stopped magnetic
monopoles. During this time an integrated luminosity of $62\pbarnt$ was
delivered. Since the binding energy of magnetic monopoles is expected to be
large in the material 
of the pipe (aluminium at that time) they should remain permanently trapped
provided that they are stable. The beam pipe was cut into long thin strips
which were each passed through a superconducting coil coupled to a
Superconducting Quantum Mechanical Interference Device (SQUID) which was sensitive down to 0.1
Dirac magnetic charges ($0.1 g_D$). Trapped
magnetic monopoles in a strip will cause a persistent current to be induced in
the superconducting coil by the magnetic field of the monopole, after complete
passage of the strip through the coil. In contrast, the induced currents from
the magnetic fields of the ubiquitous permanent magnetic dipole moments in the
material, which can be pictured as a series of equal and opposite magnetic
charges, cancel so that the current due to dipoles returns to zero after
passage of the strip.

\begin{figure}[hhh]
\vspace{-1cm}
  \begin{center}
   {\epsfig{file=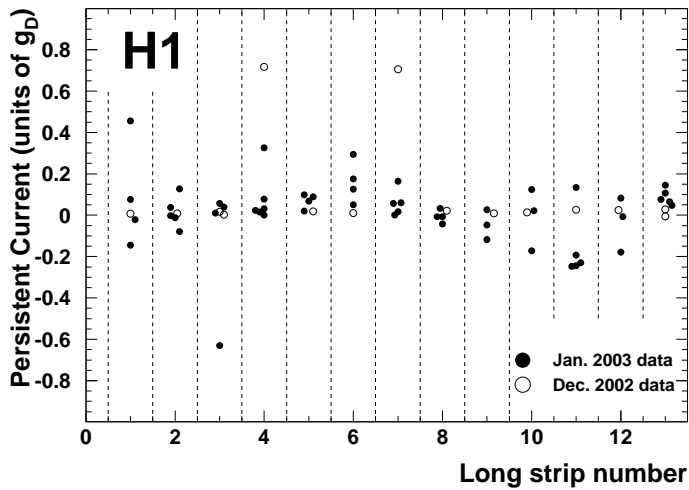,width=10cm}}
  \end{center}
%  \vspace{7.3cm}
%%
  \flushleft
  \caption{
The measured persistent current (in units of $g_D$) in each strip, after
passage through the magnetometer, plotted against strip number for 13 strips
of the central beam pipe. Some of the strip numbers are offset for clarity. It
can be seen that none of the fluctuations observed in single readings occurred
consistently in other readings on the same strip showing that no trapped
monopole was present. 
}
  \label{fig:monopoles}
\end{figure}

The values for the persistent current were converted to Dirac Monopole units
($g_D$) and are shown for different strips in Fig.~\ref{fig:monopoles}. In the
dataset of December 2002 (open circles) two of the strips measured showed
persistent currents of value expected from the passage of a magnetic charge of
about $+0.7 g_D$.  All the strips were then remeasured several times in the
set of data in January 2003, shown as closed circles in
Fig.~\ref{fig:monopoles}. None of them (except a single reading for strip 3)
showed a persistent current after traversal through the magnetometer. It was
therefore assumed that the observed persistent currents during the first set
had been caused by random jumps in the base line of the magnetometer
electronics. It can be seen from Fig.~\ref{fig:monopoles} that none of the strips
showed a persistent current which appeared consistently in more than one reading.
Thus, no magnetic monopoles were observed and charge and mass dependent
upper limits on the $e^+p$ production cross section are set.

%%%%%%%%%%%%%%%%%%begin%%figure%%%%%%%%%%%%%%%%%%%%%%%%%%%%%%%%%%%%%%%
%%%%%%%%%%%%%%%%%%%%%%%%%%%%%%end%%figure%%%%%%%%%%%%%%%%%%%%%%%%%%%%%

\section{General Search}\label{sec:general}

The H1 collaboration has performed a general search for new phenomena by
looking for deviations from the SM prediction at high transverse
momentum~\cite{general}. For the first time all event topologies involving
objects like electrons ($e$), photons ($\gamma$), muons ($\mu$), neutrinos
($\nu=$ missing particles) and jets ($j$) are investigated in a single
analysis. Event classes are defined consisting of at least two clearly
identified and isolated objects $i$ with a minimum transverse momentum of
$p_T^i > 20 \gev$. Events were found in 22 classes. The event yields span
several orders of magnitude. Overall good agreement with the SM prediction was
found. In this dataset a few events were observed in the $e j \nu$
channel where the expectation, again dominated by single $W$
production, is low.

\begin{figure}[hhh]
\vspace{-4.5cm}
  \begin{center}
\begin{picture}(0,300)
         \put(0,-15){\epsfig{file=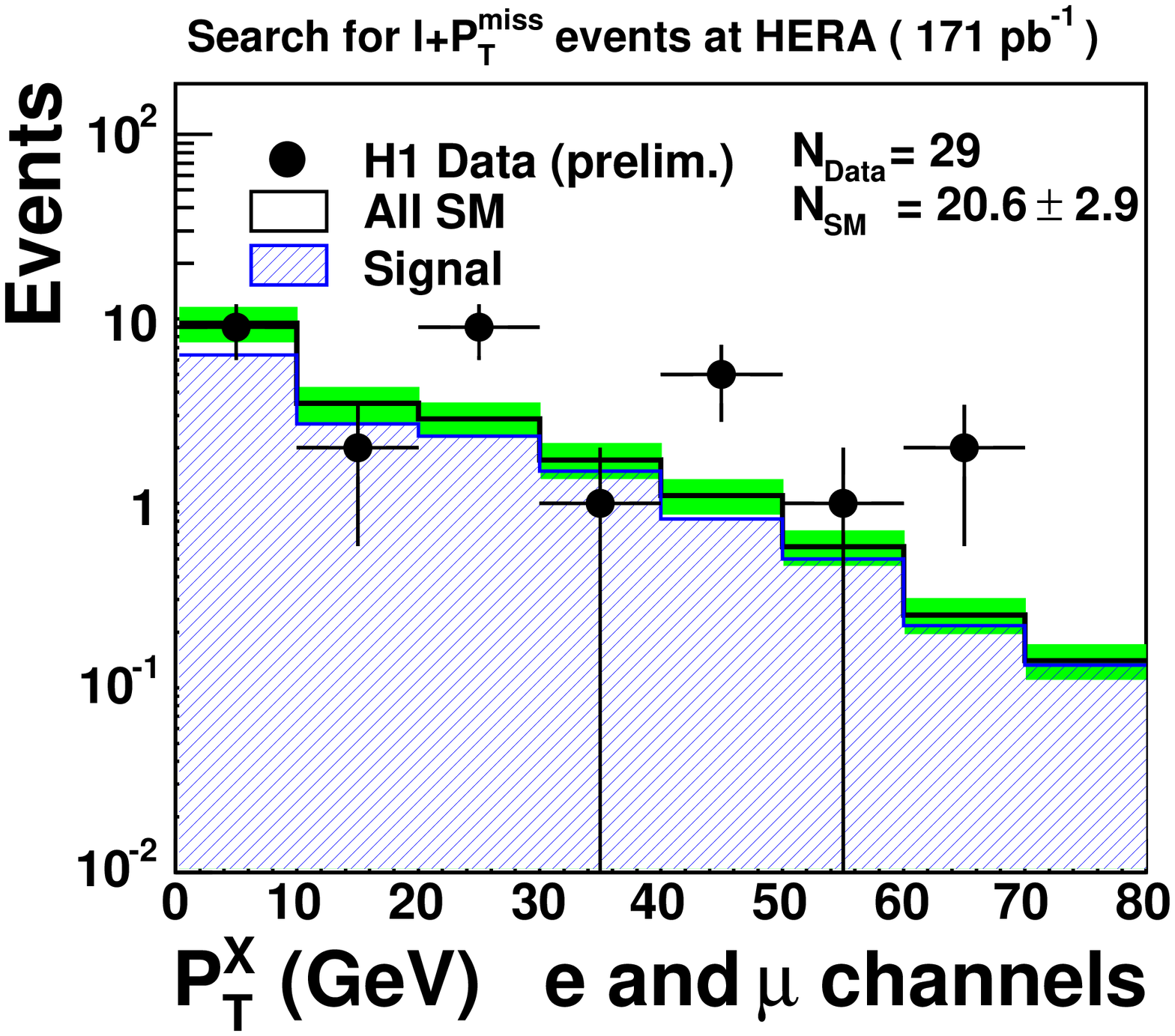,width=8cm}}
         \put(-235,-6.5){\epsfig{file=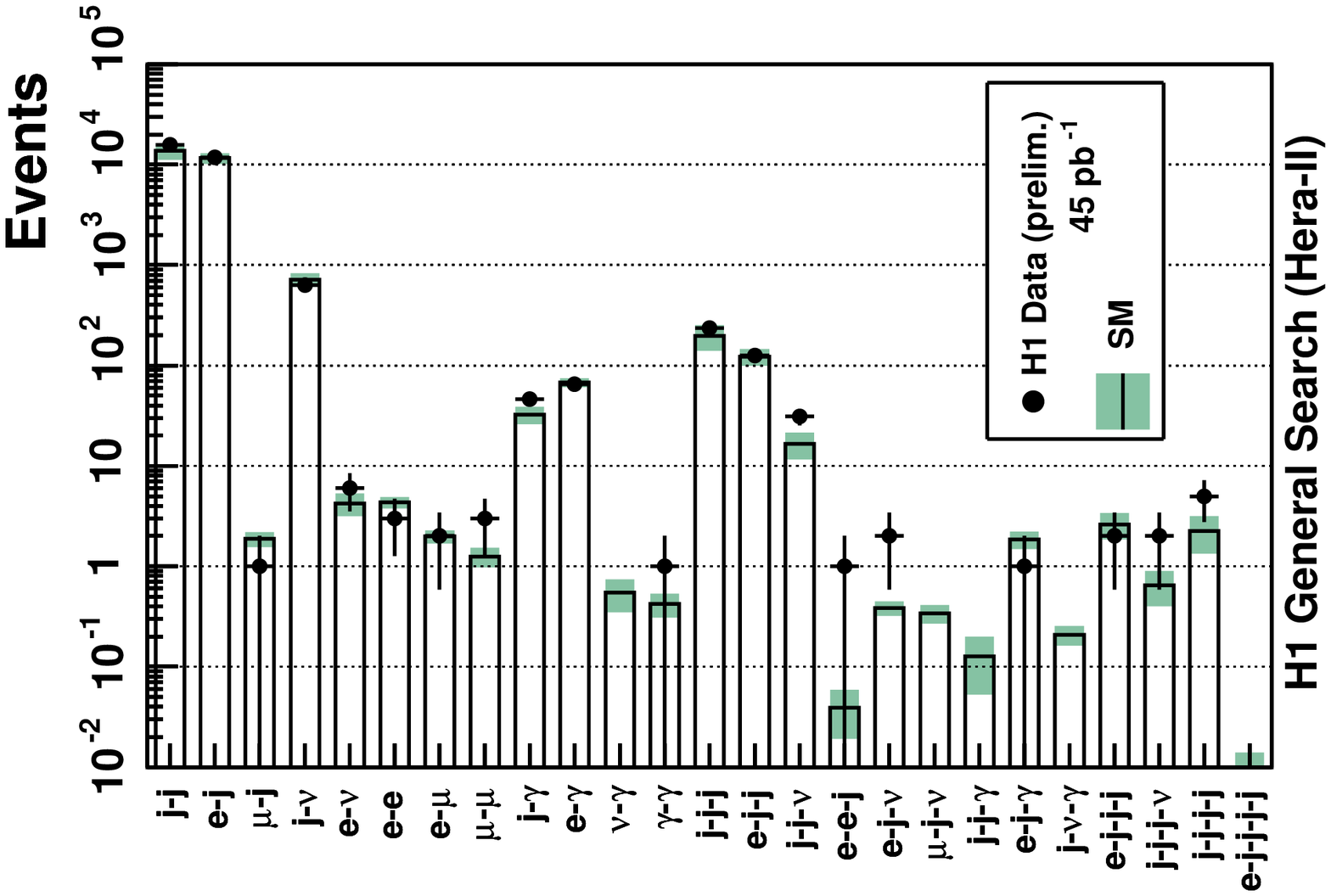,width=8.4cm}}
\end{picture}
  \end{center}
%  \vspace{8.cm}
%%
  \caption{Left: HERA II event yields in the general search for new phenomena
    at high $p_T$. Data (points) are compared to the SM prediction (histogram
    bars). The uncertainty of the SM prediction is indicated by the shaded
    band. Right: The hadronic transverse momentum distribution in the electron
    and muon channels combined compared to the SM expectation (open
    histogram). The SM expectation is dominated by real $W$ production
    (hatched histogram). The total error on the SM expectation is given by
    the shaded band.
}
  \label{fig:isogen}
\end{figure}

The same analysis was repeated with the recent data taken by the H1 experiment
in 2003/2004 based on an integrated luminosity of $45 \pbarnt$. Again good
overall agreement between data and the SM expectation was found, see
Fig.~\ref{fig:isogen} (left). In this dataset the most significant deviation
was 
found in the $ej\nu$ channel, to which also single $W$ production mainly
contributes in the SM. The excesses observed in the $ej\nu$ and $\mu j \nu$
classes are further detailed in the next section.

%
%%%%%%%%%%%%%%%%%%begin%%figure%%%%%%%%%%%%%%%%%%%%%%%%%%%%%%%%%%%%%%%
%%%%%%%%%%%%%%%%%%%%%%%%%%%%%%end%%figure%%%%%%%%%%%%%%%%%%%%%%%%%%%%%

\section{Isolated leptons with missing transverse momentum}\label{sec:isolep}

%%%%%%%%%%%%%%%%%%begin%%table%%%%%%%%%%%%%%%%%%%%%%%%%%%%%%%%%%%%%%%%
\begin{table}[hhh]
\vspace{-0.3cm}
\caption{HERA I event yields in the search for isolated leptons with missing
  transverse momentum. The numbers are given for the electron, muon and tau
  channel for different cuts $p_T^X$.\label{tab:hera1}}
\vspace{0.4cm}
\begin{center}
\begin{tabular}{|c|c|c|c|}
\hline
HERA I 1994-2000 & \mco{3}{|c|}{observed/expected} \\
\hline \hline
{\bf H1} ${\cal L}(e^\pm p)=118 \pbarnt$ & Electron & Muon & Tau \\
\hline
Full sample & 11 / $11.5 \pm 1.5$ & 8 / $2.94 \pm 0.50$ & 5 / $5.81 \pm 1.36$ \\
$p_T^X > 25 \gev$ & {\bf 5}  / $1.76 \pm 0.30$ & {\bf 6} / $1.68 \pm 0.30$ & 0 / $0.53 \pm 0.10$ \\
$p_T^X > 40 \gev$ & 3  / $0.66 \pm 0.13$ & 3 / $0.64 \pm 0.14$ & 0 / $0.22 \pm 0.05$ \\
\hline
{\bf ZEUS} ${\cal L}(e^\pm p)=130 \pbarnt$ & Electron & Muon & Tau \\
\hline
Full sample & 24 / $20.6 \pm 3.2$ & 12 / $11.9 \pm 0.6$ & 3 / $0.4 \pm 0.12$ \\
$p_T^X > 25 \gev$ & 2  / $2.9^{+ 0.59} _{-0.32}$ & 5 / $2.75 \pm 0.21$ & {\bf 2} / $0.2 \pm 0.05$ \\
$p_T^X > 40 \gev$ & 0  / $0.94 \pm 0.11$ & 0 / $0.95^{+0.14} _{-0.10}$ & 1 / $0.07 \pm 0.02$ \\
\hline
\end{tabular}
\end{center}
\vspace{-0.2cm}
\end{table}
%%%%%%%%%%%%%%%%%%end%%table%%%%%%%%%%%%%%%%%%%%%%%%%%%%%%%%%%%%%%%%%%  
%
%%%%%%%%%%%%%%%%%%begin%%table%%%%%%%%%%%%%%%%%%%%%%%%%%%%%%%%%%%%%%%%
\begin{table}[hhh]
\vspace{-0.3cm}
\caption{HERA II event yields in the search for isolated leptons with missing
  transverse momentum. The numbers are given for the electron and muon channel
  and after combination for different cuts $p_T^X$.\label{tab:hera2}}
\vspace{0.4cm}
\begin{center}
\begin{tabular}{|c|c|c|c|}
\hline
HERA II 2003-2004 & \mco{3}{|c|}{observed/expected} \\
\hline \hline
{\bf H1} ${\cal L}(e^\pm p)=53 \pbarnt$ & Electron & Muon & Combined \\
\hline
Full sample & 9 / $4.75 \pm 0.76$ & 1 / $1.33 \pm 0.19$ & 10 / $6.08 \pm 0.92$ \\
$p_T^X > 25 \gev$ & {\bf 5}  / $0.84 \pm 0.19$ & 0 / $0.85 \pm 0.13$ & {\bf 5} / $1.69 \pm 0.28$ \\
\hline
\end{tabular}
\end{center}
\vspace{-0.1cm}
\end{table}
%%%%%%%%%%%%%%%%%%%%end%%table%%%%%%%%%%%%%%%%%%%%%%%%%%%%%%%%%%%%%%%%

Events with isolated leptons and missing transverse momentum were selected by
requiring an isolated high $p_T$ lepton (electron, muon or tau) and missing
transverse momentum. For the remaining hadronic final state the
transverse momentum ($p_T^X$) is measured. In the radial plane, it is
required that the hadronic final state and the lepton are not
back-to-back, which reduces 
genuine background from deep
inelastic scattering, and ensures that the missing transverse momentum is due
to an invisible particle ($\nu$). The genuine SM ``background'' process is
single $W$ production with a leptonic decay. For the SM prediction of this
process next-to-leading order QCD 
corrections are implemented to the leading order MC generator~\cite{EPVEC} by
a reweighting method~\cite{wnlo}. 
Searches have been performed by both experiments H1 and ZEUS in the decay
channels into electrons and muons~\cite{isoleph1,isolepzeus1}, and
taus~\cite{isoleph1tau,isolepzeus2}. 

In the HERA I analyses good agreement between data and the SM expectation was
found at both experiments, see Tab.~\ref{tab:hera1}. However, after applying a
cut $p_T^X > 25 \gev$, 11 
electron and muon events were observed by the H1 experiment compared to an
expectation of $3.44 \pm 0.59$, and 2 events were observed in the tau channel
by the ZEUS collaboration compared to an expectation of $0.20 \pm
0.05$. Neither of these excesses were confirmed by the partner
experiment.

To enhance the limited statistics isolated lepton events were investigated in
the recent HERA II dataset by the H1 collaboration~\cite{highpt}. The analyses
were repeated in the electron and muon channel. In total 10 new events were
found compared to an expectation of $6.08 \pm 0.92$ events, see
Tab.~\ref{tab:hera2}. For $p_T^X > 25
\gev$, 5 events --- all found in the electron channel --- survived compared to
an expectation of $0.84 \pm 0.19$ in the electron and $0.85 \pm 0.13$ in the
muon channel.

The distribution of the transverse momentum of the hadronic final state of all
H1 data combined corresponding to an integrated luminosity of $171 \pbarnt$ is
shown in Fig.~\ref{fig:isogen} (right). For $p_T^X > 25 \gev$ a clear excess
of 16 events compared to $5.1 \pm 1.0$ expected is visible. However, more data
and more detailed studies are required to resolve unambiguously the
differences of event yields observed so far by the two collaborations H1 and
ZEUS.

\section{Conclusions}

Results on searches for leptoquarks, light gravitinos in \rpv\ SUSY
models and magnetic monopoles using the HERA I (1994-2000) $e^\pm p$ data have
been presented. No deviation from the SM prediction was found. Since that
period of data taking HERA
has seen a major upgrade program to provide 
higher luminosities and longitudinally polarised $e^\pm$ beams and first data
have been taken. Recent results including HERA II (2003/2004) $e^+ p$ data on
a general model independent search and on a search for isolated lepton events
with missing transverse momentum have been discussed. The trend of H1 at HERA
I detecting more isolated lepton events than predicted seems also to continue
at HERA II in the electron channel. At present, given the available statistics
of isolated lepton events collected at the H1 and ZEUS experiments, the
results are inconclusive and clearly more data, which are going to be
collected in the upcoming years, will help to solve the puzzle.

\end{document}